    \documentclass[aip,rsi,amsmath,amssymb,reprint]{revtex4-1}
\usepackage{graphicx}
\usepackage{dcolumn}
\usepackage{bm}
\usepackage{amsmath}
\usepackage{amssymb}
\usepackage{color}
\usepackage{subfigure}
\newcommand{\dd}{\text{d}}
\newcommand{\nn}{\nonumber\\}
\def\bal#1\eal{\begin{align}#1\end{align}}

\newcommand{\con}{{\text{cont}}}

\newcommand{\eeta}{\eta}
\newcommand{\on}{{(1)}}
\newcommand{\thr}{{(3)}}

\begin{document}

\title{Vapor-liquid equilibrium and equation of state of two-dimensional fluids
from a discrete perturbation theory}

\author{V\'ictor M. Trejos}
\affiliation{Instituto de Qu\'imica, Universidad Nacional Aut\'onoma de M\'exico,
Apdo. Postal 70213, Coyoac\'an, 04510 Ciudad de M\'exico, Mexico.}

\author{Andr\'es Santos}
\affiliation{Departamento de F\'isica and Instituto de Computaci\'on Cient\'ifica Avanzada (ICCAEx),
Universidad de Extremadura, E-06006 Badajoz, Spain}

\author{ Francisco G\'amez}
\affiliation{Department of Physical, Chemical and Natural Systems, Universidad Pablo de Olavide, 41013
Seville, Spain}
\email{fgammar@gmail.com}

\begin{abstract}

The interest in the description of the properties of fluids of restricted dimensionality is
growing for theoretical and practical reasons. In this work, we have firstly developed an analytical
expression for the Helmholtz free energy of the two-dimensional square-well fluid in the Barker--Henderson
framework. This equation of state is based on an approximate analytical radial distribution function for
$d$-dimensional hard-sphere fluids (1$\le d\le$3) and is validated against existing and new simulation results.
The so-obtained equation of state is implemented in a discrete perturbation theory able to account for general potential
shapes. The prototypical Lennard-Jones and Yukawa fluids are tested in its two-dimensional version against
available and new simulation data with semiquantitative agreement.

\end{abstract}
\maketitle

\section{Introduction}
\label{sec1}

The holy grail in the theory of liquids is perhaps the development and application, from a purely molecular perspective, of equations of state (EoSs).\cite{HM06,S13,S16} This interplay between molecular theories of
fluids and experiments with practical applications was born in the
early work of van der Waals.\cite{vanderWaals1893}

Statistical mechanics perturbation theories (PTs) for molecular or atomic
fluids\cite{Barker1967,Barker1967v2,Chandler1970,Henderson1970,Weeks1971,Solana2009,S13}
are of pivotal importance and constitute a handable tool for
the applied sciences, given their negligible computational
cost in comparison with simulations and the mathematical simplicity and intuitivity in comparison with integral
equations or density functional theory. Since for isotropic
fluids the structure is led by the repulsive
part of the global interaction, PTs are based on a decoupling of the total intermolecular potential in
perturbative and reference contributions. The reference part accounts for the structure correlation functions
of the fluid and consequently it is desirable to know it analytically.
Hence, for some simple two-dimensional (2D) models and their mixtures, the hard-disk (HD) model emerges as the most
common candidate for being used as reference in PT.
Very often, the perturbative contribution contains the
pairwise atomic interaction describing the so-called van
der Waals forces. When electrostatic forces are negligible
and the fluid is of atomic character, the functionality of
the potential energy surface becomes only dependent on
the intermolecular distance. This is the case of many
emblematic potential models, such as the widely studied hard-sphere, square-well (SW), Yukawa,
Lennard-Jones (LJ), and their 2D counterparts.\cite{Phillips1981,Reddy1986,Rovere1990, Lowen1992,Mulero1999,WenZe2011,Khordad2012,Murillo2003,Lang2014,Vaulina2014,Kryuchkov2017} Although an extensive work on three-dimensional (3D) systems can be found,\cite{Solana2009}
few works on the application of PTs to 2D models have been reported in the
literature.\cite{Mishra1984,Henderson1977}
For instance, the statistical associating fluid theory of
variable range (SAFT-VR) in its 2D version has been applied to
predict the adsorption isotherms of pure fluids and mixtures onto different kinds of surfaces.\cite{buenrostro2004,Martinez2007,Lupita2008,Castro2009,Victor2013,Victor2017,Castro2011,2013Alex,Victor2018}

Another route is the discrete PT (DPT). From theoretical grounds, the Zwanzig PT
(later popularized by Barker and Henderson\cite{Barker1967,Barker1967v2}) is the basis for the development
of the DPT\cite{Benavides1999} for atomic models \cite{Chapela2010} and
its extension to polar\cite{Benavides2011} and molecular fluids\cite{Francisco2013,Francisco2014} constitute a powerful alternative to traditional PTs.
It has been established as a versatile generator of EoSs that are analytical in terms of density,
temperature, and intermolecular parameters characterizing the interaction, so that all thermodynamic properties
can be easily obtained using common thermodynamic relations.  Although, by construction, these theories use
the hard-sphere model as a reference potential, they also provide good approximations for the thermodynamic properties
of soft-core potentials, as is the case of the LJ or Kihara fluids. Very recently, the situation
has been partly circumvented with the implementation of a temperature-dependent Barker--Henderson (BH)
diameter\cite{Cervantes2015} for the hard-core wall taking into account the repulsive van der Waals volume.
Hence, in general, the DPT can be applied to any intermolecular potential model that can be expressed as
the sum of a sequence of square-shoulder (SS) and SW steps. The strategy followed in the present work is to incorporate the approximate analytical radial distribution function (RDF) for 2D
fluids described in Refs.\ \onlinecite{Santos1993,Santos2016} into a BH high-temperature expansion for the SW2D
model. The improvement over the existing EoS for SW2D fluids motivated us to integrate this
new EoS in a general DPT able to describe arbitrary potential functions.
Both the 2D LJ and Yuwaka potentials of variable range are considered as benchmark examples.

This paper is organized as follows. In Sec.\ \ref{sec2} a description of the BH PT
for SW2D fluids is presented. A possible parameterization of the so-obtained EoS is also included for practical reasons.
A short review of the DPT procedure to map a continuous potential onto a sequence of
discrete steps is also incorporated in Sec.\ \ref{sec2}, together with a description of our Gibbs ensemble Monte Carlo (GEMC) simulation results.
Illustrative examples of the applicability of our treatment are given in Sec.\ \ref{results}. Finally, we close the paper with
Sec.\ \ref{sec-conclusion}, where the main conclusions and perspectives of this work are presented.

\section{Theoretical framework}
\label{sec2}
\subsection{Perturbation theory for two-dimensional square-well fluids}

In the framework of the  BH PT,\cite{Barker1967,Barker1967v2}
the pair potential, $ u(r)$, is decomposed into two contributions,
\begin{equation}
\label{ur}
 u(r) = u^{\text{HD}}(r) +  w(r),
\end{equation}
where $u^{\text{HD}}(r)$ and $w(r)$ are the repulsive
HD and (attractive) perturbation pair potential interactions, respectively.
The HD contribution is given by
\begin{equation}
u^{\text{HD}}(r) =
\begin{cases}
\infty, &r < \sigma,\\
0,&r \geq  \sigma ,
\end{cases}
\end{equation}
where $r$ is the distance between centers of two particles and $\sigma$ is the hard-core diameter.
For a system of particles of diameter $\sigma$ interacting via a SW pair potential, the perturbation contribution is
given by
\begin{equation}
\label{EqSW}
w^{\text{SW}}(r)=
 \begin{cases}
0,&r < \sigma,\\
\varepsilon,&\sigma \leq r < \lambda \sigma,  \\
0,&r \geq \lambda \sigma,
\end{cases}
\end{equation}
where $-\varepsilon>0$ and $\lambda\sigma$ are the depth and range of the SW potential,
respectively. If $\varepsilon>0$, Eq.\ \eqref{EqSW} describes as well the case of a (repulsive) SS interaction potential

Once the pair potential is split as described above, the corresponding Helmholtz free energy per particle, $a$,
for a general 2D fluid can be written in the BH approach as\cite{Barker1967,Barker1967v2}
\begin{equation}\label{Ageneral}
{a} = {a^{\text{id}}} + {a_\text{ex}^{\text{HD}}} + \beta  {a_1}   + \beta^2{a_2},
\end{equation}
where $\beta=1/kT$ ($k$ and $T$ being the Boltzmann constant and the temperature, respectively),
$a^{\text{id}}$ is the Helmholtz free energy of a 2D ideal gas,
$a_\text{ex}^{\text{HD}}$ is the \emph{excess} Helmholtz free energy of the reference HD fluid, and $a_1$, $a_2$ are the
first- and second-order perturbation terms, respectively.\cite{Barker1967}
The ideal gas contribution is given by
\begin{equation}
\label{Eq:aid}
\beta{a^{\text{id}}}  = \ln(\rho \lambda_B^2) -1,
\end{equation}
where $\lambda_B$ is the de Broglie wavelength and $\rho$ is the number of particles per unit area.
According to the accurate EoS for HDs proposed by Henderson,\cite{Henderson1975}
\begin{equation}\label{Phi1}
\beta{a_\text{ex}^{\text{HD}}}  = \frac{9}{8}\frac{\eta}{(1-\eta)}-\frac{7}{8}\ln(1-\eta),
\end{equation}
where $\eta=\pi \rho \sigma^2/4$ is the 2D packing fraction.

The first-order contribution to the Helmholtz free energy, $a_1$, can be written as
\begin{equation}\label{EqA1}
 \beta{a_1}  = \pi \rho  \int_{\sigma}^{\infty} dr\, w(r) r {g}^{\text{HD}}(r),
\end{equation}
where $g^{\text{HD}}(r)$ is the RDF of the HD fluid.
The second-order perturbation term is given in the so-called
``local compressibility'' approximation  by
\begin{equation}\label{EqA2}
 \beta{a_2}  = -\frac{\pi}{2}
  K^{\text{HD}}
 \rho
 \frac{\partial}{\partial \rho} \left[  \rho
 \int_{\sigma}^{\infty} dr\, w^2(r) r{g}^{\text{HD}}(r)
 \right],
\end{equation}
where  $K^{\text{HD}}=kT (\partial \rho  /\partial P^{\text{HD}})_T$ is the (reduced) isothermal compressibility of the
HD fluid, which is given by
\begin{equation}
 K^{\text{HD}}(\eta) = \frac{(1-\eta)^3}{1+\eta+\frac{3}{8}\eta^2-\frac{1}{8}\eta^3},
\end{equation}
according to the Henderson EoS.\cite{Henderson1975}

Equations \eqref{EqA1} and \eqref{EqA2} apply to any perturbation contribution $w(r)$. In the special case of the SW interaction, Eq.\ \eqref{EqSW}, the perturbation terms $a_1$ and $a_2$ can be rewritten as
\begin{subequations}
\label{Eq:a1&a2}
 \begin{equation}
\beta a_1^{\text{SW}}=\varepsilon a_1^*,\quad \beta a_2^{\text{SW}}=\varepsilon^2 a_2^*,
 \end{equation}
\begin{equation}
\label{Eq:a1}
a_1^*(\eta,\lambda)=4\eta  J(\eta,\lambda),
\end{equation}
\begin{equation}
\label{Eq:a2}
a_2^*(\eta,\lambda)   = -\frac{1}{2} K^{\text{HD}}(\eta)\eta
\frac{\partial a_1^*(\eta,\lambda)}{\partial\eta},
\end{equation}
\end{subequations}
where we have introduced the dimensionless quantity
\begin{equation}\label{EqJ2}
 J(\eta,\lambda) \equiv \sigma^{-2} \int_{\sigma}^{\lambda \sigma}dr\, r{g}^{\text{HD}}(r,\eta).
\end{equation}

The integral \eqref{EqJ2} requires the prior knowledge of the RDF for a HD fluid. Obviously, it is highly desirable to have an  analytical approximation for ${g}^{\text{HD}}(r)$, thermodynamically consistent with the Henderson EoS, in such a way that the integral $J(\eta,\lambda)$ can be obtained for arbitrary values of both $\eta$ and $\lambda$. This would allow, by insertion of Eqs.\ \eqref{Eq:aid}, \eqref{Phi1}, and \eqref{Eq:a1&a2} into Eq.\ \eqref{Ageneral}, to determine the free energy as an explicit function of density, temperature, and the two parameters ($\varepsilon$ and $\lambda$) characterizing the SW potential. Here, we will make use of the approximation for ${g}^{\text{HD}}(r)$ proposed in Ref.\ \onlinecite{Santos1993}, and recently generalized to any dimensionality $1\leq d\leq3$,\cite{Santos2016} to obtain $J(\eta,\lambda)$ analytically. The details are presented in the Appendix.


\subsection{Effective packing fraction and its parameterization}
\label{sec2B}

Following the methodology used for 3D and 2D
fluids,\cite{Alejandro1997,Patel2005,Lupita2008} one can apply the mean-value theorem to express the integral \eqref{EqJ2} as
\begin{equation}
\label{eff1}
J(\eta,\lambda)=g^{\text{HD}}(\xi,\eta) \frac{\lambda^2 -1 }{2},
\end{equation}
where $\xi(\eta,\lambda)$ is a certain appropriate distance.
As discussed in Gil-Villegas {\it et al.},\cite{Alejandro1997} one can define an \emph{effective} packing fraction $\eta_{\text{eff}}(\eta,\lambda)$ such that
\begin{equation}
\label{eff2}
  g^{\text{HD}}(\xi,\eta) =  g_\con^{\text{HD}}(\eta_{\text{eff}}),
\end{equation}
where $g_\con^{\text{HD}}(\eta)$ is the  RDF at contact. Its expression consistent with the Henderson EoS\cite{Henderson1975} is
\begin{equation}
\label{eff3}
g_\con^{\text{HD}} (\eta)=\frac{1-7\eta/16}{(1-\eta)^2}.
\end{equation}
Combination of Eqs.\ \eqref{eff1}--\eqref{eff3} yields a quadratic equation for $\eta_{\text{eff}}$ whose physical solution is
\begin{equation}
\eta_{\text{eff}}(\eta,\lambda)=1-\frac{1+\sqrt{1+18Q(\eta,\lambda)/7}}{Q(\eta,\lambda)},
\end{equation}
where $Q(\eta,\lambda)\equiv \frac{64}{7}J(\eta,\lambda)/(\lambda^2-1)$.
Therefore, Eq.\ \eqref{Eq:a1} can be rewritten as
\begin{equation}
\label{Eq:a1_alt}
\beta a_1^*(\eta,\lambda)=2(\lambda^2-1)\eta g_\con^{\text{HD}} (\eta_{\text{eff}}(\eta,\lambda)).
\end{equation}

Following  Patel {\it et al.},\cite{Patel2005} we have obtained a parameterization for $\eta_{\text{eff}}$ (by taking numerical results for
$8\times 10^{-4}\le \eta\le 0.63$ and $1.02 \le\lambda \le12$) in the form
\begin{equation}\label{etaeff}
 \eta_{\text{eff}}(\eta,\lambda) = \eta\frac{ c_1(\lambda)  + c_2(\lambda) \eta }{ [1+c_3(\lambda) \eta]^3},
\end{equation}
where the coefficients $c_{1,2,3}$ are given in matrix form by
\begin{align}
\label{cmatrix}
\left(
\begin{aligned}
    c_1 \\
    c_2 \\
    c_3 \\
\end{aligned}
 \right)
= &
\begin{pmatrix}
          0.15605  &       -0.60341     &     4.10347     &    -2.33312\\
         -0.82505   &       12.03157     &    -40.40351    &      33.23906\\
          9.73879   &      -47.09168      &    66.35256     &    -28.17232\\
\end{pmatrix}
\nonumber \\ &
\cdot
\begin{pmatrix}
   1/\lambda  \\
  1/\lambda^2  \\
  1/\lambda^3 \\
  1/\lambda^4 \\
 \end{pmatrix}.
\end{align}
Use of the (approximate) parameterization \eqref{etaeff} in Eq.\ \eqref{Eq:a1_alt}, together with Eq.\ \eqref{Eq:a2}, provides manageable analytical expressions for $a_1^*$ and $a_2^*$ that  are convenient from the point of view of practical applications.

\subsection{Discrete perturbation theory}

As commented above, the DPT is a BH PT capable of generating
EoSs analytical in the parameters that characterize the intermolecular interactions.\cite{Benavides1999}
It can be applied to a great variety of intermolecular potential models that can be discretized as the
sum of $p$ steps of SS and SW potentials of variable energy scale $\{\varepsilon_i\}$ and width $\{\lambda_i\}$.
The construction of the DPT is based on the use of the hard-core model as reference potential, what results in a good
approximations for the thermodynamic properties of low-density properties of soft-core potentials (such as the
LJ fluid) but fails in the prediction of the high and intermediate part of the phase diagram
(see, for example, the 3D LJ case treated in Refs.\ \onlinecite{Benavides2011,Chapela2010}).

A $p$-step discrete potential has the form
     \begin{equation}
\label{a1}
u^{\text{dis}}(r)=
\begin{cases}
\infty  ,& r<\sigma, \\
\varepsilon_1  ,& \sigma<r<\lambda_1\sigma , \\
\varepsilon_2  ,& \lambda_1\sigma<r<\lambda_2\sigma , \\
\vdots&\vdots \\
\varepsilon_p  ,& \lambda_{p-1}\sigma<r<\lambda_p\sigma , \\
0,&r>\lambda_p.
\end{cases}
\end{equation}
If $u^{\text{dis}}(r)$ is intended to represent a given continuous potential $u(r)$, it is convenient to choose
\begin{subequations}
\label{udis_u}
\begin{equation}
 \varepsilon_i=u\left(\sigma\frac{\lambda_{i-1}+\lambda_i}{2}\right),\quad \lambda_0=1,
 \end{equation}
 \begin{equation}
 \lambda_i=1+i \Delta \lambda,\quad \Delta \lambda=\frac{\lambda_p-1}{p}.
\end{equation}
\end{subequations}

By using the notation $u^{\text{SW}}(r;\varepsilon,\lambda)$ for the SW perturbation function \eqref{EqSW}, the perturbation contribution to $u^{\text{dis}}(r)$ can be written as\cite{Benavides1999}
\begin{equation}
  w^{\text{dis}}(r)=\sum_{i=1}^p  \left[w^{\text{SW}}(r;\varepsilon_i,\lambda_i)-w^{\text{SW}}(r;\varepsilon_i,\lambda_{i-1})\right].
\end{equation}
A similar expression holds for $[w^{\text{dis}}(r)]^2$. As a consequence, from Eqs.\ \eqref{EqA1} and \eqref{EqA2}, one obtains the following forms for the first- and second-order contributions of the free energy corresponding to the potential \eqref{a1}:
\begin{subequations}
\label{e2}
\begin{equation}
\beta a_1^{\text{dis}}=\sum_{i=1}^p \varepsilon_i\left[a_1^*(\eta,\lambda_i)-a_1^*(\eta,\lambda_{i-1})\right],
\end{equation}
\begin{equation}
\beta a_2^{\text{dis}}=\sum_{i=1}^p \varepsilon_i^2\left[a_2^*(\eta,\lambda_i)-a_2^*(\eta,\lambda_{i-1})\right],
\end{equation}
\end{subequations}
where, in the 2D case, $a_1^*(\eta,\lambda)$ and $a_2^*(\eta,\lambda)$ are given by Eqs.\ \eqref{Eq:a1} and \eqref{Eq:a2}, respectively.
In summary, the Helmholtz free energy per particle corresponding to the interaction potential \eqref{a1} is
\begin{align}
\beta a^{\text{dis}}=&\beta a^{\text{id}}+\beta a^{\text{HD}}+\sum_{i=1}^p \beta\varepsilon_i\left[a_1^*(\eta,\lambda_i)-a_1^*(\eta,\lambda_{i-1})\right]\nonumber\\
&+\sum_{i=1}^p
(\beta\varepsilon_i)^2\left[a_2^*(\eta,\lambda_i)-a_2^*(\eta,\lambda_{i-1})\right].
\end{align}

Once the Helmholtz free energy is known as a function of density, temperature, and the set of potential parameters, the pressure ($P$) and chemical potential ($\mu$) can be obtained by standard thermodynamic relations as
\begin{equation}
Z\equiv \frac{P}{\rho kT}=\rho\frac{\partial \beta a}{\partial\rho},\quad \beta\mu=\beta a+Z.
\end{equation}
The vapor-liquid phase boundaries are obtained by solving the nonlinear set of equations that establish the
conditions of thermal, mechanical, and chemical equilibrium.
Those conditions will be fulfilled by equating the temperature ($T^v=T^{l}$),
the pressure ($P^v=P^{l}$), and the chemical potential ($\mu^v=\mu^{l}$) of both vapor ($v$) and liquid ($l$) phases.

\subsection{Gibbs ensemble Monte Carlo simulations}

In this work, we studied the vapor-liquid phase coexistence of SW2D and LJ2D fluids
by using  the GEMC simulation method proposed by Panagiotopolous.\cite{Panagio1}
We initially placed $N=1152$ particles uniformly distributed in two boxes with equal areas and number of particles
$N_1=N_2=576$ with $N=N_{1}+N_{1}$. The following MC movements were performed:
(i) displacements of particles,
(ii) interchange of particles between the two boxes, and
(iii) exchange of area keeping constant the
total area of the system {i.e.,} $S=S_1+S_2$.
We define a one MC cycle as the action of performing randomly the aforementioned
operations at the ratio of $9\times 10^4:1:100$.
Equilibration required $5\times 10^5$ MC cycles, followed by
$4\times 10^5$ MC cycles for average production. The acceptance ratio of particles displacements and
area exchange were fixed to $40\%$.
In both boxes, periodic boundary conditions and minimum-image
convention in the two Cartesian coordinates were assumed.

\section{Results}
\label{results}

\subsection{Vapor-liquid equilibrium of two-dimensional square-well fluids}

\begin{figure}
  \subfigure{\label{fig1a}
\hspace*{-1.5cm}\includegraphics[width=0.75\columnwidth]{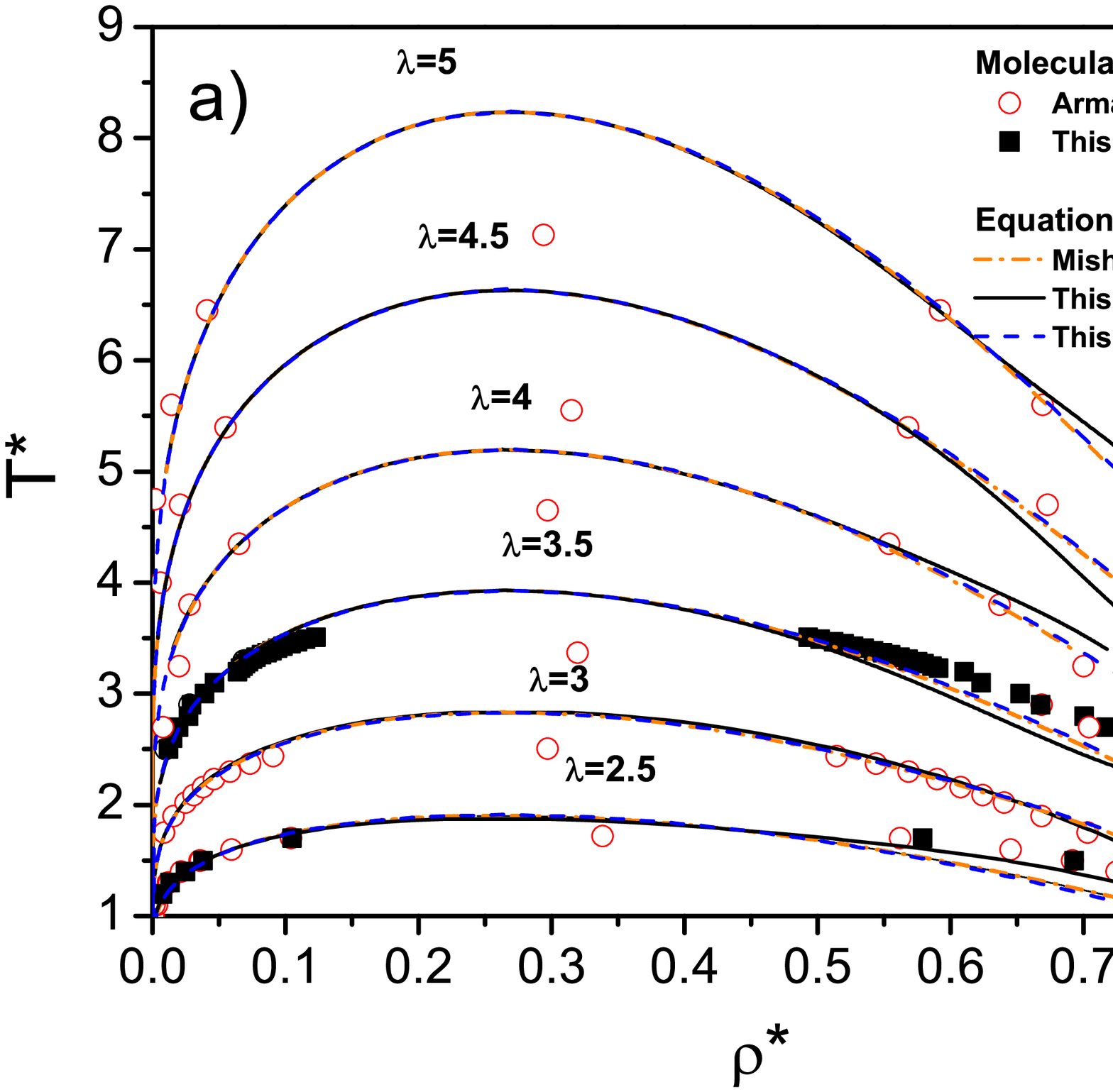}}
\subfigure{\label{fig1b}
\hspace*{-1.5cm}\includegraphics[width=0.75\columnwidth]{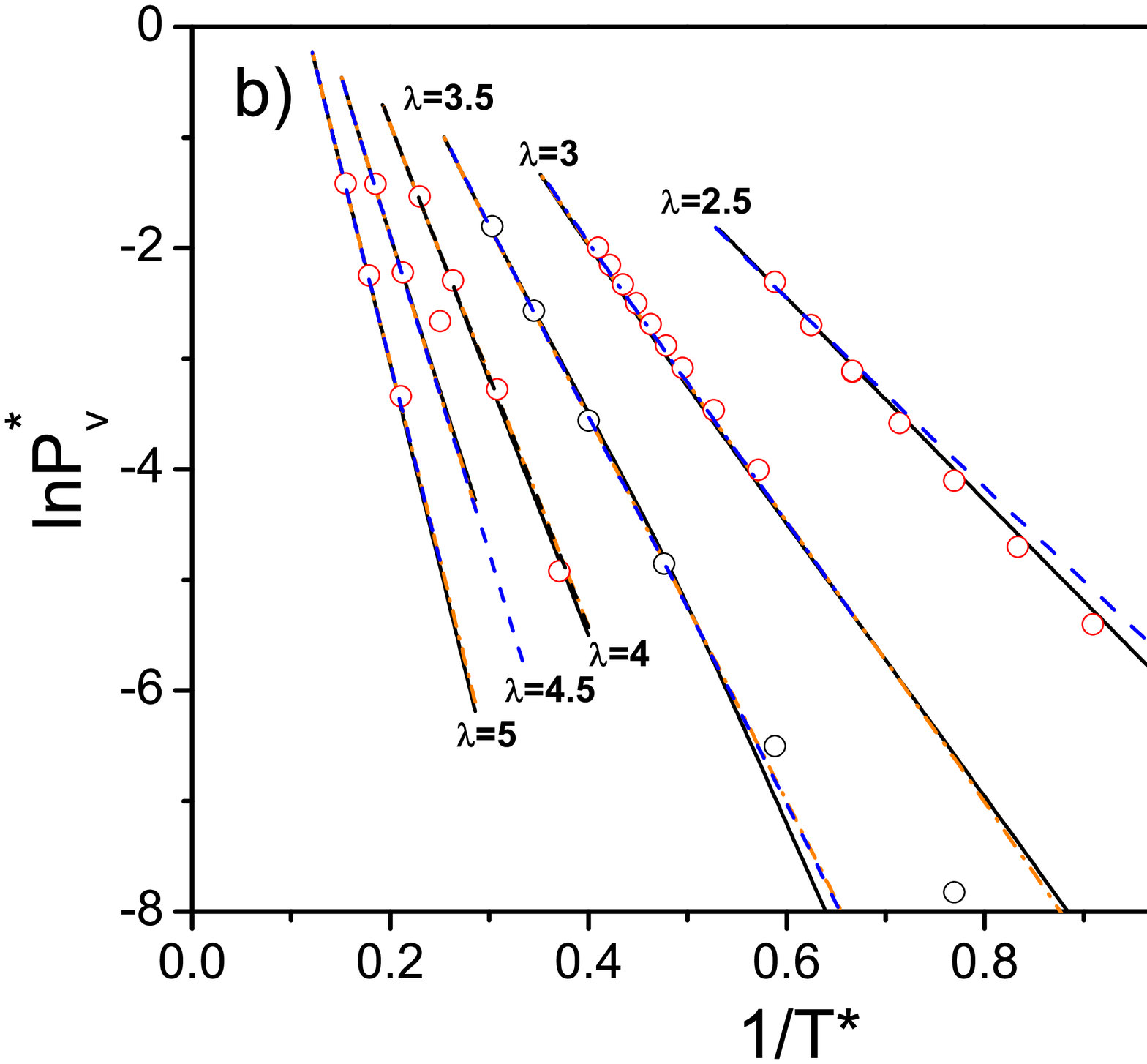}}
\caption{VLE of the SW2D fluid with $\lambda\geq 2.5$. Symbols correspond to the simulation results
of Armas {\it et al.}\cite{Armas2013} (open circles) and our new simulation results (full squares). Dash-dotted lines
corresponds to the EoS reported in Ref.\ \onlinecite{Mishra1984}, while the solid and  dashed lines
are the predictions of the full (solid) and fitted (dashed) EoS described in this work. Here, $T^*=kT/|\varepsilon|$, $\rho^*=\rho\sigma^2$, and $P^*=P\sigma^2/|\varepsilon|$.}
\label{Fig1}
\end{figure}

\begin{figure}[ht]
\subfigure{\label{fig2a}
 \hspace*{-1.5cm}\includegraphics[width=0.75\columnwidth]{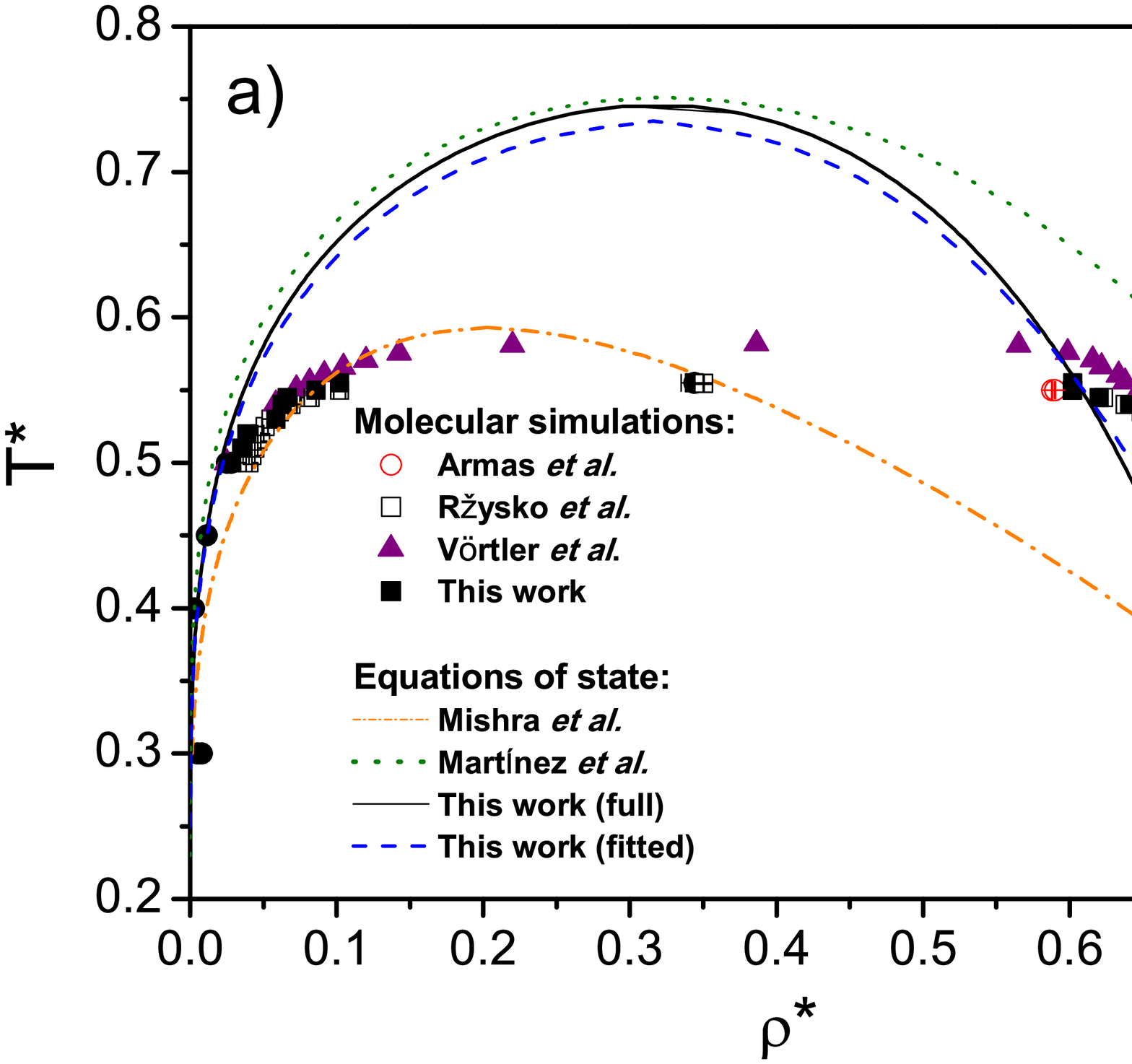}}
  \subfigure{\label{fig2b}
\hspace*{-1.5cm}\includegraphics[width=0.75\columnwidth]{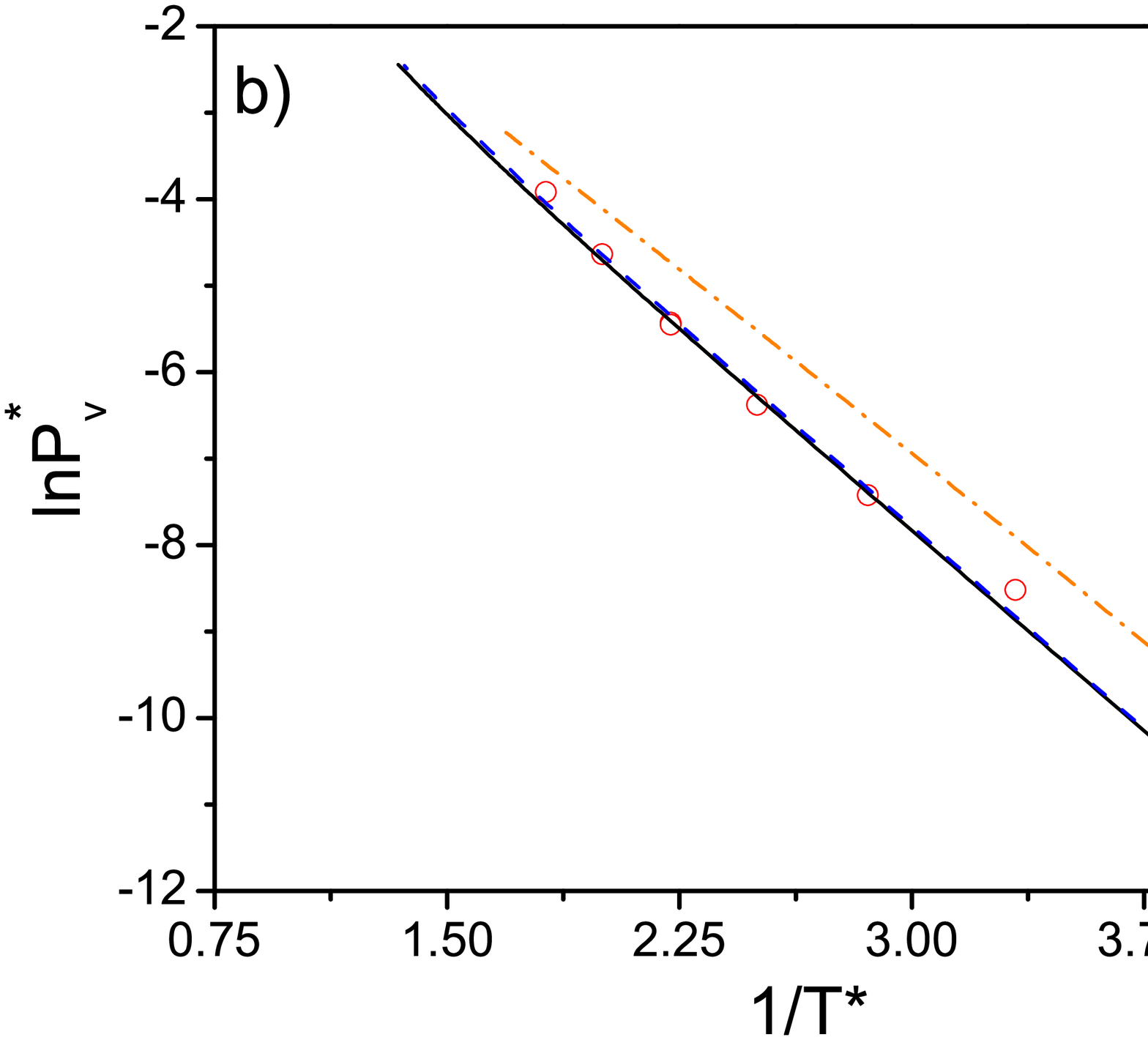}}
\caption{VLE of the SW2D fluid with $\lambda$=1.5. Symbols correspond to the simulation results
of Armas {\it et al.}\cite{Armas2013} (open circles),  R{\.z}ysko \emph{et al.}\cite{Rzysco2010} (open squares), V{\"o}rtler \emph{et al.}\cite{Vortler2008} (full triangles), and our new simulation results (full squares).
The dash-dotted and dotted lines corresponds to the EoSs reported in Refs.\ \onlinecite{Mishra1984,Martinez2007}, respectively, while the solid and  dashed lines are the predictions of
the full (solid) and fitted (dashed) EoS described in this work.}
\label{Fig2}
\end{figure}

The vapor-liquid equilibrium (VLE) of SW2D fluids is established here as a severe test of
the accuracy of a given EoS. We have separated our results into two figures. In Fig.\ \ref{Fig1},
the VLE is shown for SW2D fluids of long ranges. Our simulation and theoretical results are very close to the simulation results of
Armas {\it et al.}\cite{Armas2013} and to the EoSs of Mishra and Sinha\cite{Mishra1984} and
Mart\'inez {\it et al.}\cite{Martinez2007} The approximation of Mishra and Sinha consists in ignoring the fluid structure of the RDF at long distances, i.e., $\delta J(\lambda) \equiv-\sigma^{-2}\int_{\lambda \sigma }^{\infty }dr\, r \left [  g^{\text{HD}}(r) -1 \right]\approx 0$.
This assumption deteriorates the accuracy of the associated EoS for shorter ranges, as can be observed
in Fig.\ \ref{Fig2}, which includes additional simulations results of R{\.z}ysko \emph{et al.}\cite{Rzysco2010} and V{\"o}rtler \emph{et al.}\cite{Vortler2008} The reason behind this behavior can be found in Fig.\ \ref{Fig:J} of the Appendix, where it can be checked that $\delta J$
differs notably from zero at short ranges. It should be noted that the oscillatory behavior of $\delta J$ introduces
a nontrivial systematic error that depends strongly on $\lambda$. For $\lambda=1.5$, the EoS of Mishra and Sinha dramatically
underestimates the liquid density. In contrast, this liquid branch is overestimated by the SAFT approach.\cite{Martinez2007}
The results of the EoS presented in this paper reproduces the VLE with semi-quantitative agreement,
while the saturation pressure is accurately described. Remarkably, the apparent density anomaly observed
in simulations at low temperatures is also predicted within our approach qualitatively.
Similar results have been obtained with the parameterized EoS described in Sec.\ \ref{sec2B}.
It should be pointed out that the accuracy of PTs worsens as the dimensionality decreases.
Moreover, in 2D fluids, the VLE exhibits a high flatness near the critical point that cannot be reproduced by a PT.
In all cases, the results obtained with the parameterization described by Eqs.\ \eqref{Eq:a1_alt}--\eqref{cmatrix}
are similar to those obtained with the full calculations employing Eqs.\ \eqref{Eq:a1&a2}.

\subsection{Vapor-liquid equilibrium and equation of state of two-dimensional Lennard-Jones fluids}

\begin{figure}[ht]
  \subfigure{\label{fig3a}
\hspace*{-1.5cm}\includegraphics[width=0.75\columnwidth]{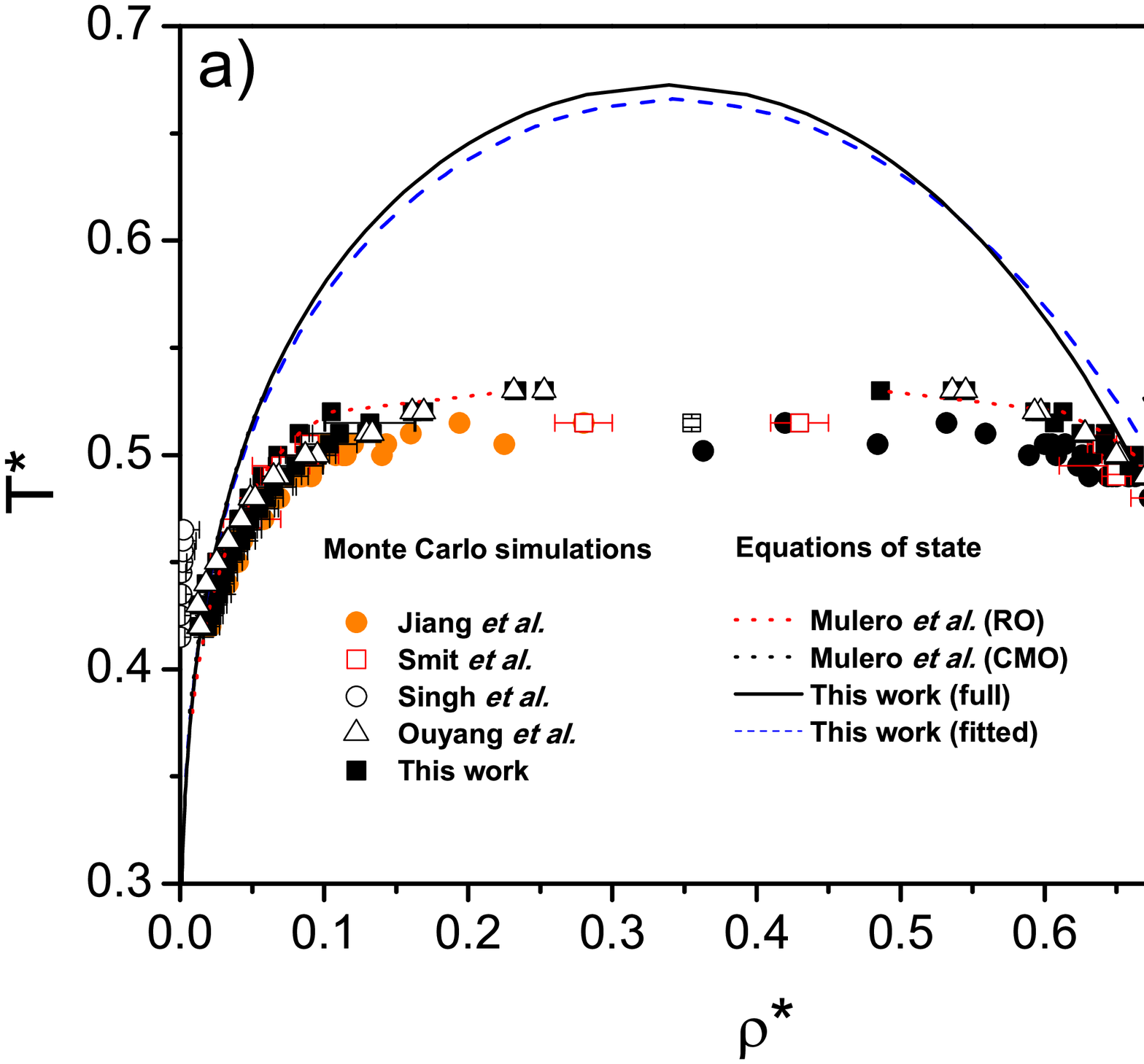}}
  \subfigure{\label{fig3b}
\hspace*{-1.5cm}\includegraphics[width=0.75\columnwidth]{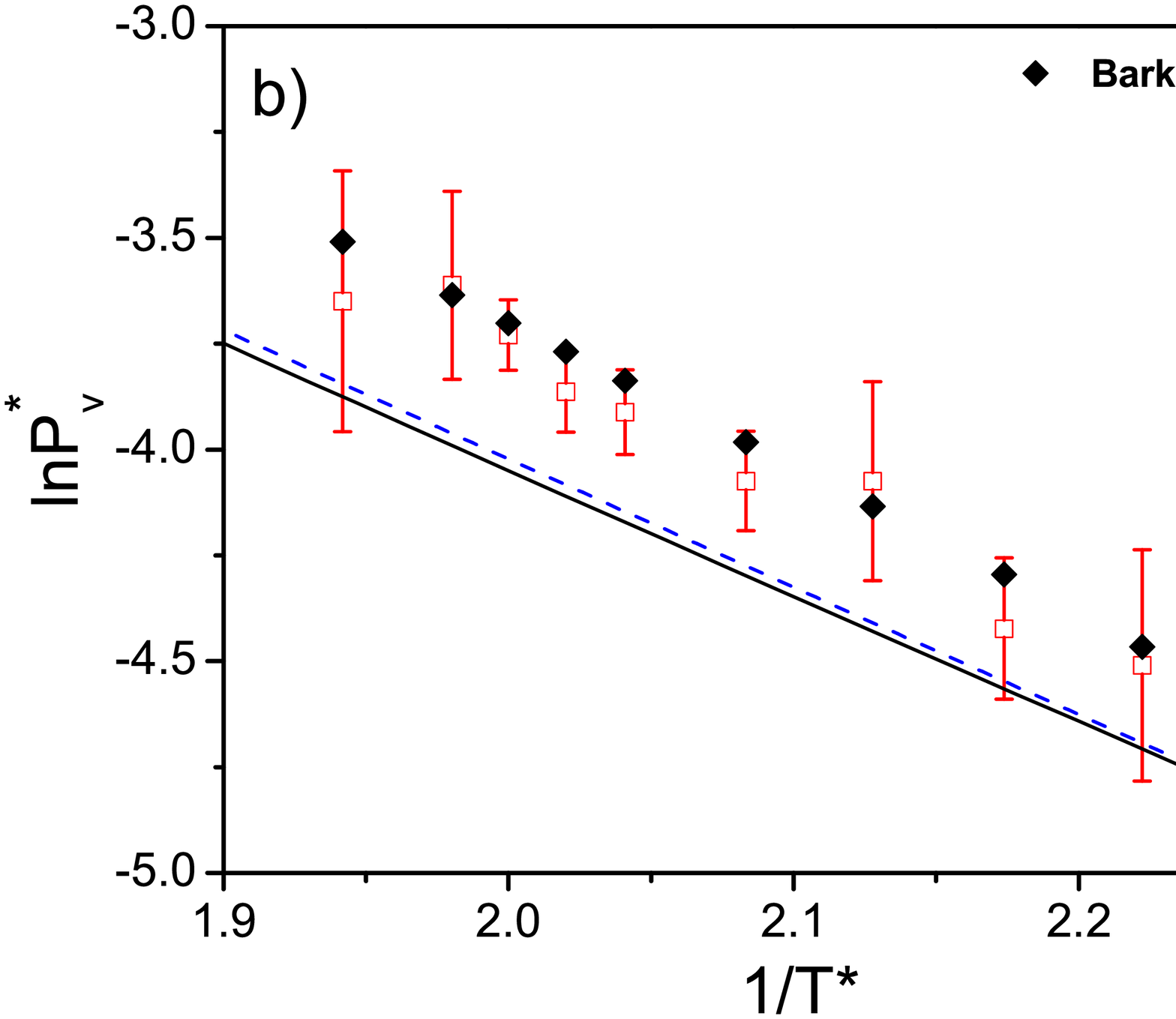}}
\caption{VLE of the  LJ2D fluid. Symbols correspond to the simulation results
from Refs.\ \onlinecite{Jiang1995,Smith1987,dePablo1990,WenZe2011}, the parameterization of Ref.\ \onlinecite{Abraham1981}, and the new GEMC results obtained in this work, as labeled in the graph. The solid and  dashed lines
are the predictions of the full (solid) and fitted (dashed) EoSs described in this work and the semiempirical expresions (denoted as CMO and RO for the Cuadros--Mulero and Reddy--O'Shea EoSs, respectively) of Ref.\ \onlinecite{Mulero1999}}.
\label{Fig3}
\end{figure}

\begin{figure}[ht]
  \subfigure{\label{fig4a}
\hspace*{-1.5cm}\includegraphics[width=0.75\columnwidth]{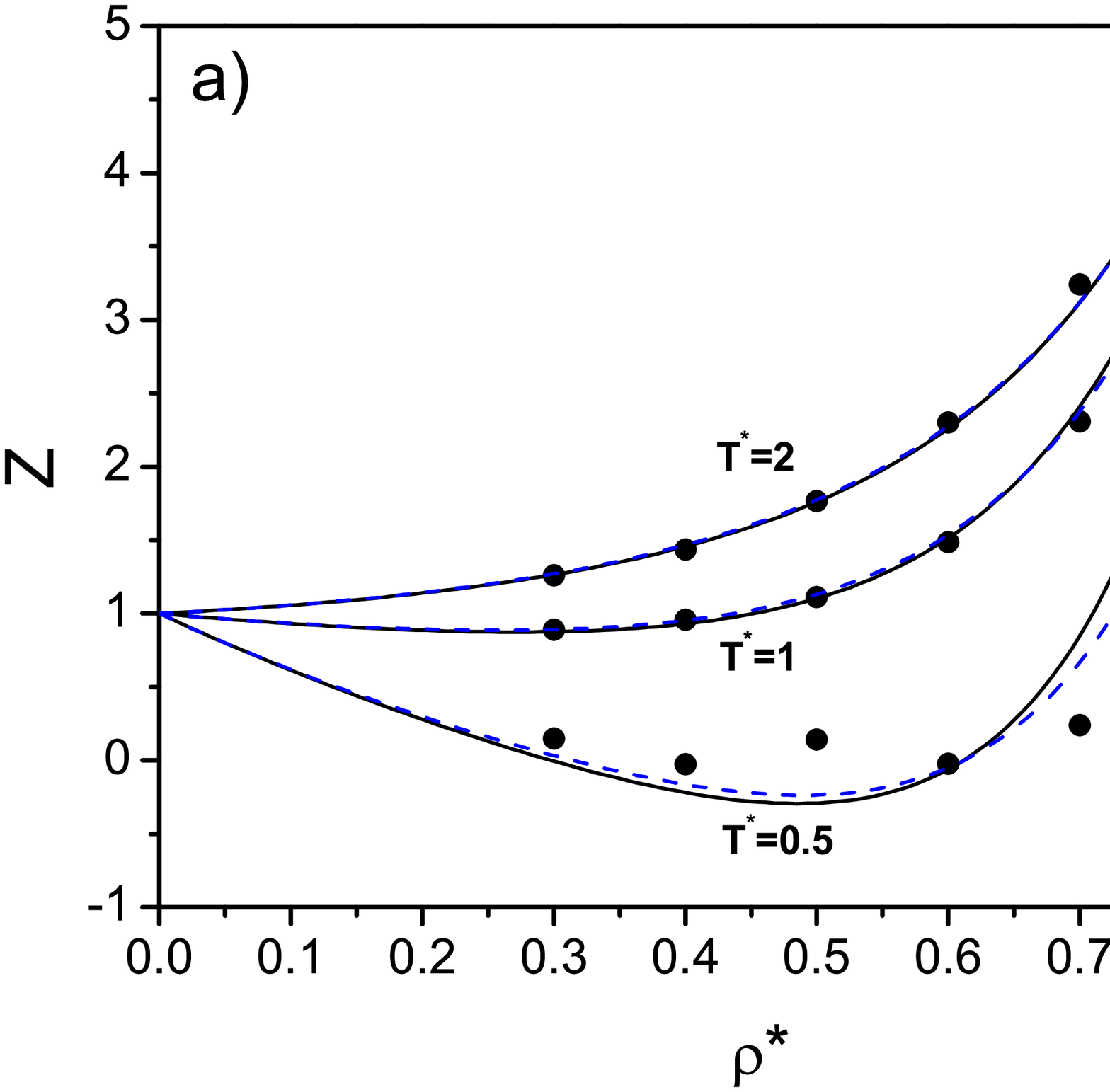}}
  \subfigure{\label{fig4b}
\hspace*{-1.5cm}\includegraphics[width=0.75\columnwidth]{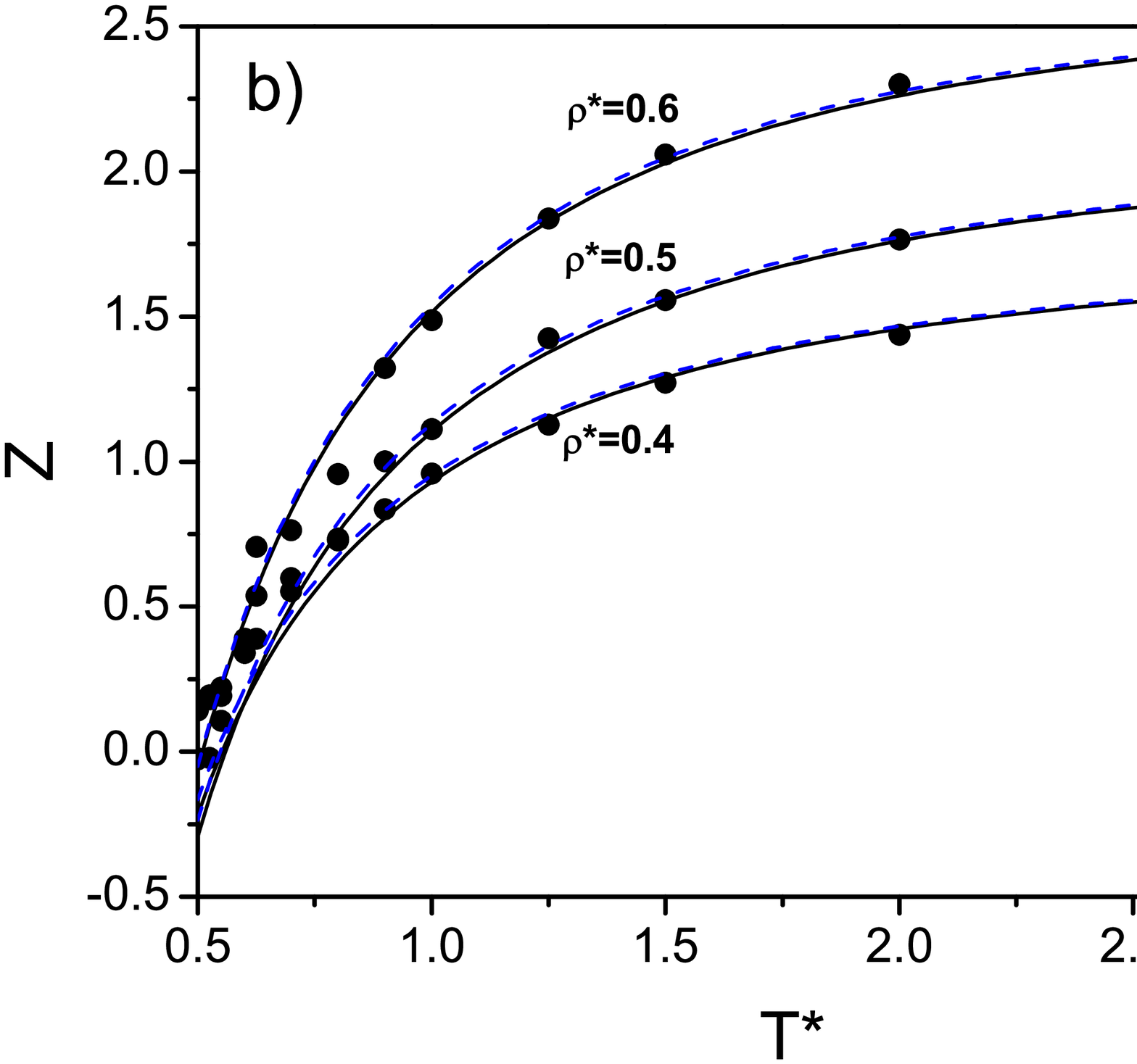}}
\caption{EoS of the LJ2D fluid. Symbols correspond to the simulation results
from Ref.\ \onlinecite{Henderson1977}. The solid and  dashed lines are the predictions of the full (solid)
and fitted (dashed) EoSs described in this work.}
\label{Fig4}
\end{figure}

The well-known LJ pair potential is
\begin{equation}
u^{\text{LJ}}(r)=4\varepsilon\left [ \left ( \frac{\sigma}{r} \right )^{12}-\left ( \frac{\sigma}{r} \right )^{12}\right],
\end{equation}
where $\sigma$ and $\varepsilon$ are the diameter  and the  well depth, respectively. The LJ potential is a coarse-grained model extensively used in computer simulations.
As such, it has a much higher computational simplicity in comparison with those potentials coming from
quantum chemistry calculations, but by the price of being less accurate.
In essence, the LJ potential provides a simple
description of the repulsive electronic interaction due to the quantum Pauli principle, as well as of the attractive part
coming from the induction and
dispersive interactions.\cite{LJ1924}

The discrete version of  the LJ potential is given by Eqs.\ \eqref{a1} and \eqref{udis_u} with $\lambda_p=\lambda_{c}$, where $\lambda_{c}$ is a cutoff distance for the LJ tail. It can be calculated by requiring $|u^{\text{LJ}}(r)/\varepsilon|<10^{-6}$ for $r>\lambda_c\sigma$, which gives
 $\lambda_{c}\simeq 12.6$. The number of steps is chosen as $p=3$ in the repulsive part of the potential profile ($1\leq r/\sigma\leq 2^{1/6}$) and  $p=\lfloor 10({\lambda_{c}-1})\rfloor $, where $\lfloor x \rfloor$ denotes the floor function of argument $x$, in the attractive region ($r/\sigma>2^{1/6}$).

Up to here, a soft potential has been converted into a hard potential. This approximation is reasonably good for some
cases treated up to now. However, the integration of the Mayer function would help in the treatment
of the repulsive part by mapping the LJ potential onto an effective HD of diameter given in the BH approach as
\begin{equation}\label{eq1}
d(\beta)=\int_{0}^{\sigma}dr\,\left\{  1-\exp \left[ -\beta u^{\text{LJ}}(r)  \right] \right\}.
\end{equation}
In this work, we have used the parameterization of  $d(\beta)$ reported in Ref.\ \onlinecite{Prausnitz1986}.
This approach is shown to have a positive effect in reproducing the VLE and
EoS coming from simulation data of the LJ2D potential.

The resulting VLE coming from both PTs is plotted in Fig.\ \ref{Fig3} in comparison with our own simulation results and those obtained from different sources in the literature.\cite{Jiang1995,Smith1987,dePablo1990,Abraham1981,WenZe2011} A more pronounced deviation for the saturation pressure, as compared with the SW case, is observed. In fact, the underestimation of the vapor pressure when the perturbation series is truncated to second order is well established in the literature.\cite{Fischer1984}.  Besides being of semiempirical character, the EoSs in Ref.\ \onlinecite{Mulero1999} do not substantially improve the agreement of the proposed analytical EoS with the simulation results and, moreover,  we avoid the ill-behavior near the critical region observed for those semiempirical formulations.

In Figs.\ \ref{fig4a} and \ref{fig4b}, the compressibility factor $Z$ obtained by MC simulation in
Ref.\ \onlinecite{Henderson1977} as a function of $\rho^{*}$ and $T^{*}$ in the medium-temperature and
density regimes, respectively, is shown. The temperature and density regimes comprise both supercritical
and subcritical regions. The agreement of simulation results and our PTs is excellent, except for low temperature and high density, and are
of similar accuracy as the semiempirical PT proposed by Henderson.\cite{Henderson1977} The accuracy of the parameterized (fitted) EoS for the SW model is well transferred to the DPT approach, as readily observed in Figs.\ \ref{Fig3} and \ref{Fig4}.


\subsection{Vapor-liquid equilibrium of two-dimensional Yukawa fluids}

\begin{figure}[ht]
\hspace*{-1.5cm}\includegraphics[width=0.75\columnwidth]{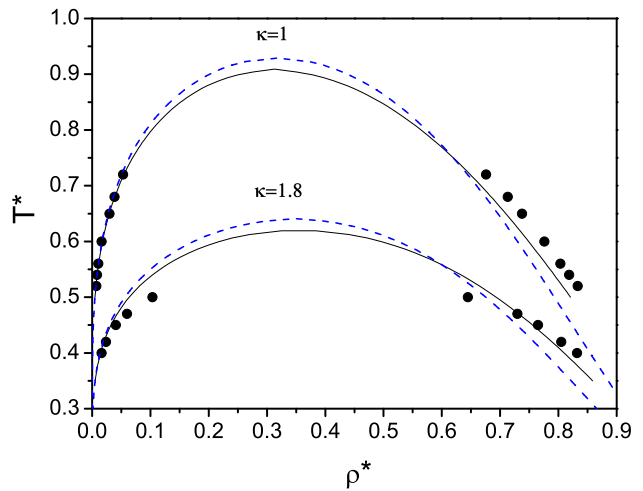}
 \caption{VLE of the 2D Yukawa fluids with $\kappa=1$ and $1.8$. Symbols correspond to
 simulation results
from Ref.\ \onlinecite{Maldonado2012}. The solid and  dashed lines are the predictions of the full (solid)
and fitted (dashed) EoSs described in this work.}
\label{Fig5}
\end{figure}

The Yukawa (or screening Coulomb) potential has become popular
nowadays in soft matter physics in the description of the electrostatic interaction between colloids and other mesoscopic
systems in the dilute limit.\cite{Hiemenz1997} Its functional form is given by
\begin{equation}
 u^{\text{Y}}(r) =
 \begin{cases}
   \infty,&r<\sigma,\\
   \displaystyle{-\varepsilon\frac{\exp[-\kappa(r/\sigma-1)]}{r/\sigma}},&r>\sigma,
 \end{cases}
\end{equation}
where $\kappa$ is the (reduced) inverse screening length.
Its discrete version can again be constructed from Eqs.\ \eqref{a1} and \eqref{udis_u}, with $p=\lfloor 10({\lambda_{c}-1})\rfloor $ and a cutoff value $\lambda_p=\lambda_c$ such that $|u^{\text{Y}}(\lambda_c\sigma)/\varepsilon|=10^{-6}$. The solution is $\lambda_c=W(10^6\kappa e^{\kappa})/\kappa$, where $W(x)$ is the Lambert function.

To the best of our knowledge, the only available results for the 2D Yukawa model can be found in Ref.\ \onlinecite{Maldonado2012},
but with a slightly different functional form, as required for the molecular dynamics method employed there. Given the peculiar results obtained in Ref.\ \onlinecite{Maldonado2012} for  values of $\kappa$ higher than $1.8$ (with densities well above $\rho^*=1$), we report  our results only for $\kappa=1$ and $1.8$ in Fig.\ \ref{Fig5}. As expected, there is a detrimental effect in the agreement as $\kappa$ increases because it causes the potential to be stiffer. This fact implies a non-optimized sequence of step functions in which some of them present
high  values of $|\varepsilon_i|$, where the high-temperature perturbation expansion no longer holds.
This effect has been checked to be crucial for large $\kappa$-values, but it propagates
for low $\kappa$-values up to the exact Kac limit $\kappa\rightarrow0$. As expected, similar conclusions can be traced when using the parameterized EoS.

\section{Conclusions}

\label{sec-conclusion}

In this paper an EoS for SW2D fluids was obtained in the framework of the high-temperature
BH PT.\cite{Barker1967,Barker1967v2} The key ingredient in the EoS is
an approximate analytical expression for the RDF of
HDs proposed in Refs.\ \onlinecite{Santos1993,Santos2016}. This equation has been tested against simulation results and other
theoretical EoSs. The EoS proposed here shows a better agreement in describing the VLE of a SW2D fluid of
middle-short range ($\lambda=1.5$) that those reported in the literature.
Besides, this EoS was implemented in an extension of the DPT, originally developed
for 3D potentials. Since the HD RDF is not restricted to any value of $\lambda$,
the discretization of the potential has \emph{a priori} no constraints on the potential range, but on the molecular packing.

We have tested the 2D DPT against simulation results for the LJ and Yukawa fluids.
Given the perturbative treatment of the attractive part of the interactions, as well as the approximate character of the RDF, our results are in semiquantitative agreement with the reported MC calculations. It seems fair to mention that this EoS constitutes a new member to be included in
the scarce list of (analytical) EoSs for 2D fluids. A common feature for PTs is that they
do not predict correctly the critical properties because critical fluctuations are neglected. This fact is
highlighted in the 2D cases. Some improvement could be achieved by incorporating those fluctuations
by the renormalization group theory, as proposed by White\cite{White1992,Salvino1992}
and successfully applied by Lovell {\it et al.}\cite{Llovell2004,Llovell2006} to the soft-SAFT EoS.

Finally, the EoS proposed here could be easily extended to incorporate molecular fluids following the assumption
given in Refs.\ \cite{Francisco2013,Francisco2014}. This extension of the theory would allow one to reproduce the thermodynamics of molecular systems (as 2D oligomers\cite{Patra2012}), would also serve as a guide for performing computer simulations of layered anisotropic fluids, or could be used as
reference in the theoretical modeling of confined liquid crystal
phases. Last but not least, given that the RDF employed here owns a general analytical solution
for an arbitrary dimension $d$, our results can also be extended to fractal
fluids.\cite{Torquato2002}

\section{Supplementary Material}

See the supplementary material for a Fortran 90 code for
the evaluation of the integral \eqref{3.3}.

\begin{acknowledgments}
A.S. acknowledges the financial support of the
Ministerio de Econom\'ia y Competitividad (Spain) through Grant No.\ FIS2016-76359-P, partially financed by ``Fondo Europeo de Desarrollo Regional'' funds.
He is also grateful to M. L\'opez de Haro for useful comments.
\end{acknowledgments}




\appendix*
\section{Analytical evaluation of the integral $J(\eta,\lambda)$}
\label{Appendix}

In this appendix, the analytical evaluation of the integral $J(\eta,\lambda)$ defined by Eq.\ (\ref{EqJ2}) is detailed. It is based on the approximate RDF for HDs proposed in Refs.\ \onlinecite{Santos1993,Santos2016}.

\subsection{Radial distribution function of hard disks}

The RDF of HDs is obtained under the assumption that the structure and spatial correlations of
HD fluids share some features with those of a hard-rod ($d=1$) and hard-sphere ($d=3$) fluids.\cite{Santos1993}
The RDF of HDs is given within this approximation as follows,\cite{Santos1993,Santos2016}
\begin{equation}\label{6}
g(r,\eeta)=\alpha(\eeta) g^{(1)}\left(r,\gamma_1(\eeta)\eeta\right)
+[1-\alpha(\eeta)]g^{(3)}\left(r,\gamma_3(\eeta)\eeta\right),
\end{equation}
where $g^{(1)}$ is the exact RDF for hard rods,\cite{S16}  $g^{(3)}$ is the RDF for hard spheres as obtained from the Percus--Yevick (PY) equation,\cite{Wertheim1963,Wertheim1964} and we have simplified the notation $ g^\text{HD}\to g$. Moreover,  the mixing parameter $\alpha$ and the scaling factors $\gamma_1$, $\gamma_3$
are the following functions of $\eta$:
\begin{subequations}
\begin{equation}
\alpha(\eeta)=\frac{H(\eeta)-H^{(3)}\left(\gamma_3(\eeta)\eeta\right)}{H^{(1)}\left(\gamma_1(\eeta)\eeta\right)-H^{(3)}\left(\gamma_3(\eeta)\eeta\right)},
\label{10}
\end{equation}
\begin{equation}
\label{8}
\gamma_1(\eeta)=\frac{g_\con(\eeta)-1}{\eeta g_\con(\eeta)},
\end{equation}
\begin{equation}
\label{9}
\gamma_3(\eeta)=\frac{1+4g_\con(\eeta)-\sqrt{1+24g_\con(\eeta)}}
{4\eeta g_\con(\eeta)}.
\end{equation}
\end{subequations}
Here, the contact values of the RDF for $d=1,2,3$ are given by
\begin{subequations}
\label{g1-1D&3D}
\begin{equation}
g^{(1)}_{\con}(\eeta)=\frac{1}{1-\eeta},
\label{g1-1D}
\end{equation}
\begin{equation}
g_\con(\eeta)=\frac{1-c\eeta}{(1-\eeta)^2},
\label{g1-2D}
\end{equation}
\begin{equation}
g^{(3)}_{\con}(\eeta)=\frac{1+\eeta/2}{(1-\eeta)^2}.
\label{g1-3D}
\end{equation}
\end{subequations}
According to the Henderson EoS,\cite{Henderson1975} $c=\frac{7}{16}=0.4375$ in Eq.\ \eqref{g1-2D}, but a better value is $c=2\sqrt{3}/\pi-2/3\simeq 0.4360$.
In Eq.\ \eqref{10}, the moment
\begin{equation}
\label{Eq:H}
H(\eeta)=-\int_0^\infty \dd r\, r [g(r,\eeta)-1]
\end{equation}
for $d=1,2,3$ is
\begin{subequations}\label{2+3}
\begin{equation}
H^{(1)}(\eeta)=\frac{1}{2}-\frac{2}{3}\eeta+\frac{1}{4}\eeta^2,
\label{2}
\end{equation}
\begin{equation}
H(\eeta)=\frac{\frac{1}{2}-\frac{1}{4}c\eeta(3-\eeta)}{1+\eeta+(1-2c)\eeta^2(3-\eeta)},
\label{4}
\end{equation}
\begin{equation}
H^{(3)}(\eeta)=\frac{\frac{1}{2}-\frac{1}{20}\eeta(2-\eeta)}{1+2\eeta}.
\label{3}
\end{equation}
\end{subequations}


\subsection{Radial distribution function of hard rods}
In the case $d=1$, the exact analytical expression for the Laplace transform
\begin{equation}
 G^{(1)}(s,\eta) = \int_0^{\infty} dr\, e^{-rs} g^{(1)}(r,\eta)
\end{equation}
is given by\cite{Heying2004,S16}
\begin{equation}
  G^{(1)}(s,\eta) = \frac{1}{\eta} \frac{e^{-s}}{1+ s(1-\eta)/\eta -e^{-s}  },
\end{equation}
where, without loss of generality, $\sigma=1$ has been taken as length unit.
By expanding $G^{(1)}(s,\eta)$  in powers of $e^{-s}$, the inverse Laplace transform
can readily be obtained term by term. Then, the RDF for hard rods $g^{(1)}(r,\eeta)$ is given by
\begin{align}\label{A3}
 g^\on(r,\eeta)=&\frac{1}{\eeta}\sum_{\ell=1}^\infty \left(\frac{\eeta}{1-\eeta}\right)^\ell\frac{(r-\ell)^{\ell-1}}{(\ell-1)!}\nn
&\times e^{-(r-\ell)\eeta/(1-\eeta)}\Theta(r-\ell),
\end{align}
where $\Theta(x)$ is the Heaviside step function of argument $x$.


\subsection{Radial distribution function of hard spheres}
The analytical solution of the PY integral equation in the three-dimensional case ($d=3$) relies upon the Laplace transform of $r g^{(3)}(r,\eta)$,
\begin{equation}
 G^{(3)}(s,\eta) = \int_0^{\infty} dr\, e^{-rs} r g^{(3)}(r,\eta).
\end{equation}
The solution is\cite{Wertheim1963,Wertheim1964,HM06,S16}
\begin{equation}
  G^{(3)}(s,\eta) = s \frac{F(s,\eta)  e^{-s}}{  1+ 12 \eta F(s,\eta) e^{-s}  },
\end{equation}
where
\begin{subequations}
\begin{equation}
F(s,\eeta)=-\frac{1}{12\eeta}\frac{1+L_1(\eeta)s}{1+S_1(\eeta)s+S_2(\eeta)s^2+S_3(\eeta)s^3},
\end{equation}
\begin{align}
L_1(\eeta)=&\frac{1+\eeta/2}{1+2\eeta},\quad S_1(\eeta)=-\frac{3}{2}\frac{\eeta}{1+2\eeta},
\\
S_2(\eeta)=&-\frac{1}{2}\frac{1-\eeta}{1+2\eeta},\quad
     S_3(\eeta)=-\frac{1}{12\eeta}\frac{(1-\eeta)^2}{1+2\eeta}.
\end{align}
\end{subequations}
As in the one-dimensional case, the function $G^{(3)}(s,\eta)$ can be expanded in powers of $e^{-s}$, thus allowing to write analytically the RDF as
\begin{equation}
g^\thr(r,\eeta)=\frac{1}{r}\sum_{\ell=1}^\infty \left(-12\eeta\right)^{\ell-1}{\Psi}_{\ell}(r-\ell,\eeta){\Theta(r-\ell)},
\label{A9}
\end{equation}
with
\begin{subequations}
\begin{equation}\label{A9b}
{\Psi}_\ell(r,\eeta)=\sum_{j=1}^\ell \frac{\sum_{i=1}^3 a_{\ell j}^{(i)}(\eeta)e^{s_i(\eeta) r}}
{(\ell-j)!(j-1)!}r^{\ell-j},
\end{equation}
\begin{equation}\label{alj}
a_{\ell j}^{(i)}(\eeta)=\lim_{s\to s_i}\left(\frac{\partial}{\partial s}
\right)^{j-1}\left\{s\left[(s-s_i)F(s,\eeta)\right]^\ell\right\},
\end{equation}
\end{subequations}
where  $s_i(\eeta)$, ($i=1,2,3$) are the three roots of the
cubic equation $1+S_1(\eeta) s+S_2(\eeta) s^2+S_3(\eeta) s^3=0$.

\subsection{Analytical evaluation of $J(\eta,\lambda)$}

\begin{figure}
  \subfigure{\label{appenA}
\hspace*{-1.5cm}\includegraphics[width=0.75\columnwidth]{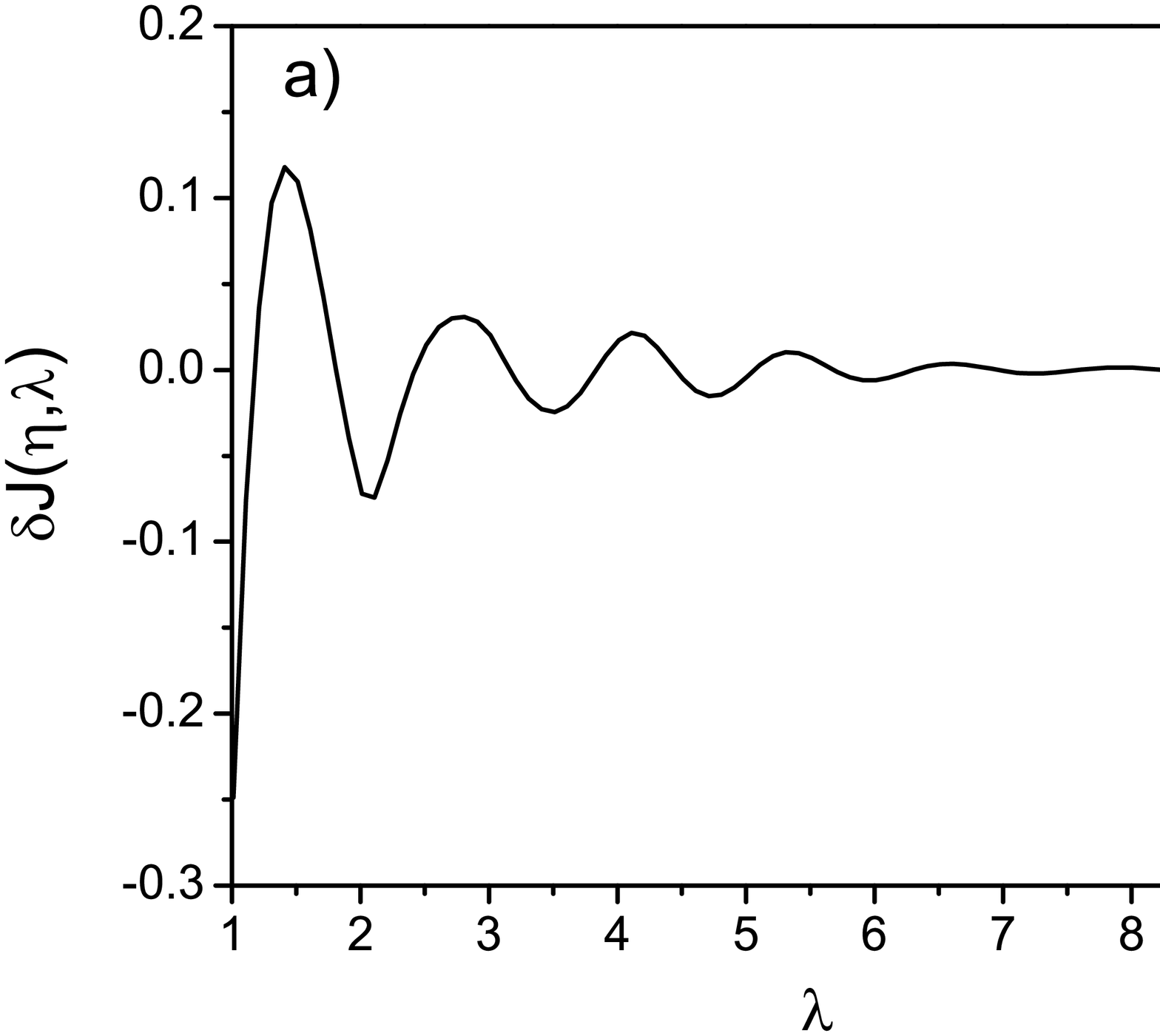}}
  \subfigure{\label{appenB}
\hspace*{-1.5cm}\includegraphics[width=0.75\columnwidth]{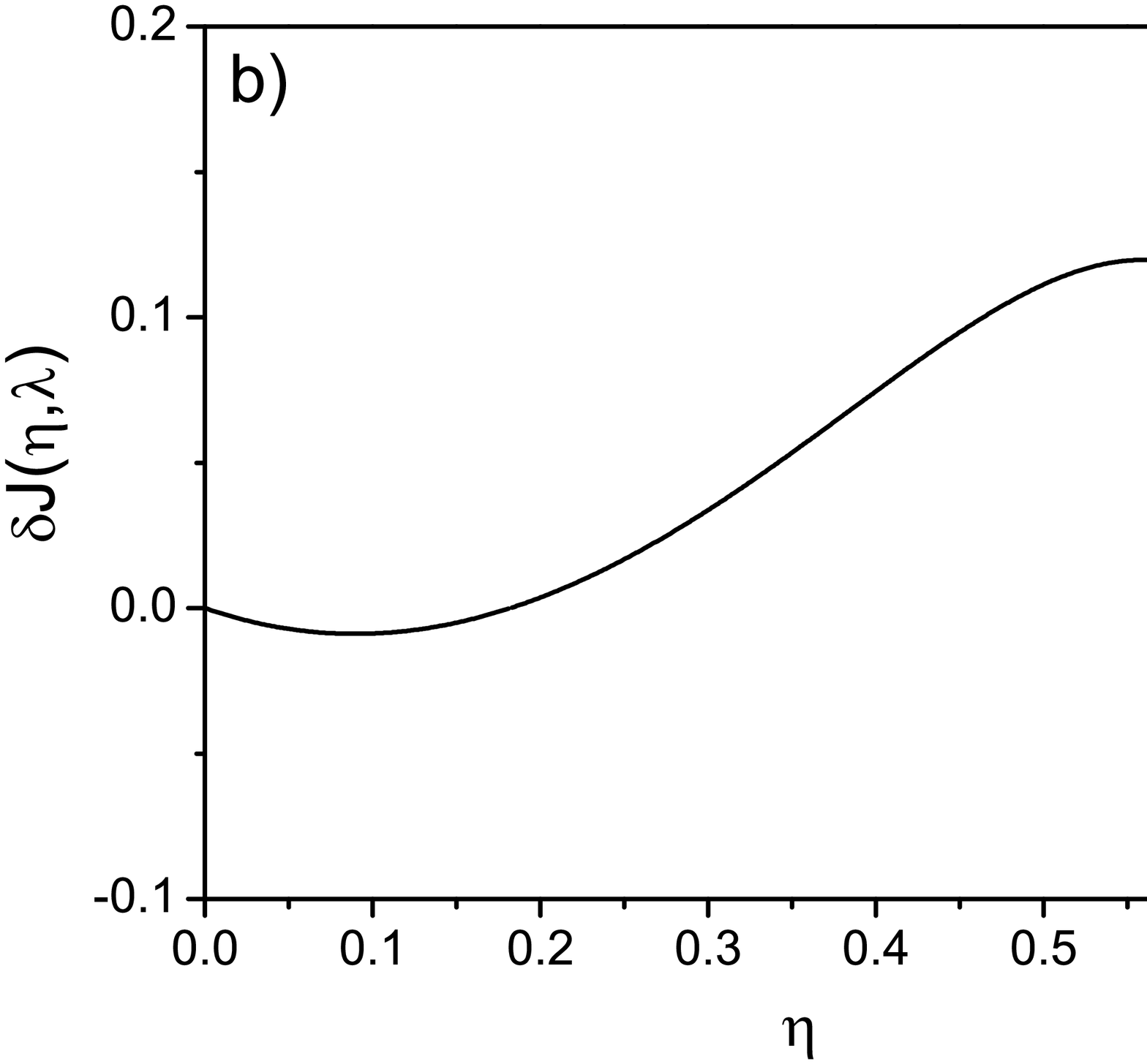}}
 \caption{Plot of $\delta J(\eeta,\lambda)=J(\eeta,\lambda)-\left[\frac{1}{2}\lambda^2-H(\eeta)\right]$  (a) as a function of $\lambda$ at $\eeta=0.5$ and (b) as a function of $\eeta$ at $\lambda=1.5$.}
 \label{Fig:J}
\end{figure}
Insertion of Eq.\ \eqref{6} into Eq.\ \eqref{EqJ2} yields
\begin{equation}\label{3.3}
 J(\eeta,\lambda)=\alpha(\eeta) J^{(1)}\left(\gamma_1(\eeta)\eeta,\lambda\right)
+[1-\alpha(\eeta)]J^{(3)}\left(\gamma_3(\eeta)\eeta,\lambda\right),
\end{equation}
where
\begin{equation}
J^{(d)}\left(\eeta,\lambda\right)=\int_1^\lambda dr\, r g^{(d)}(r,\eta).
\end{equation}
Thus, we can focus on $J^\on(\eeta,\lambda)$ and $J^\thr(\eeta,\lambda)$.

From Eq.\ \eqref{A3}, $J^{(1)}$ can be rewritten as
\begin{align}
\label{3.6}
J^\on=&\frac{1}{\eeta}\sum_{\ell=1}^{ \lfloor\lambda \rfloor}\left(\frac{\eeta}{1-\eeta}\right)^\ell\int_0^{\lambda-\ell}dx\,(x+\ell)\frac{x^{\ell-1}}{(\ell-1)!}
e^{-x\eeta/(1-\eeta)}\nn
=&\frac{1}{\eeta}\sum_{\ell=1}^{\lfloor\lambda \rfloor}\ell\left[\frac{1-\eeta}{\eeta}\Phi_\ell\left(\frac{\eeta(\lambda-\ell)}{1-\eeta}\right)+
\Phi_{\ell-1}\left(\frac{\eeta(\lambda-\ell)}{1-\eeta}\right)\right],
\end{align}
where  we have introduced the function
\begin{equation}\label{3.5}
\Phi_\ell(t) \equiv \frac{1}{\ell!}\int_0^t dx\, e^{-x} x^\ell
 = 1-e^{-t} \sum_{j=0}^\ell \frac{t^j}{j!}.
\end{equation}

Next, from Eqs.\ \eqref{A9} and \eqref{A9b}, $J^{(3)}$ is given by
\begin{align}
J^\thr=&\sum_{\ell=1}^{\lfloor\lambda \rfloor}(-12\eeta)^{\ell-1}\int_0^{\lambda-\ell} dx\,\Psi_\ell(x,\eeta)\nn
=&\sum_{\ell=1}^{\lfloor\lambda \rfloor}(-12\eeta)^{\ell-1}\sum_{j=1}^\ell \frac{1}{(j-1)!}\sum_{i=1}^3 \frac{a_{\ell j}^{(i)}}{\left(-s_i\right)^{\ell-j+1}}
\nn
&\times\Phi_{\ell-j}\left(-s_i(\lambda-\ell)\right).
\label{3.7}
\end{align}

By inserting Eqs.\ \eqref{3.6} and \eqref{3.7} into Eq.\ \eqref{3.3} we find an \emph{analytical} representation
of the integral $J(\eeta,\lambda)$. Given a value of $\lambda$, the number of terms in the sums of Eqs.\ \eqref{3.6} and \eqref{3.7} are $\lfloor\lambda\rfloor$, and $3\lfloor\lambda\rfloor(\lfloor\lambda\rfloor+1)/2$, respectively.
Thus, from a practical point of view, the above representation might present problems if $\lambda>10$.
In that case, it might be convenient to use the asymptotic behavior of $J$ for large $\lambda$. To that end, note the identity [see Eqs.\ \eqref{EqJ2} and \eqref{Eq:H}]
\begin{equation}
J(\eeta,\lambda)=\frac{1}{2}\lambda^2-H(\eeta)+\delta J(\eeta,\lambda),
\end{equation}
where
\begin{equation}
\delta J(\eeta,\lambda)\equiv-\int_\lambda^\infty dr\, r [g(r,\eeta)-1].
\end{equation}
If $\lambda$ is large, $\delta J$ can be neglected, so that
\begin{equation}\label{3.8}
J(\eeta,\lambda)\approx \frac{1}{2}\lambda^2-H(\eeta).
\end{equation}

The dependence of $\delta J(\eeta,\lambda)=J(\eeta,\lambda)-\left[\frac{1}{2}\lambda^2-H(\eeta)\right]$ as
a function of both $\lambda$ and $\eta$ is shown in Fig.\ \ref{Fig:J}. As seen in Fig.\ \ref{appenA}, Eq.\ \eqref{3.8} is an excellent approximation for $\lambda>10$. On the other hand, as expected,  $\delta J$ is non-negligible if $\lambda$ is not very far from $1$. Figure \ref{appenB} shows that $\delta J$ presents an oscillatory behavior with respect to density.

The Fortran 90 program for evaluating the function $J(\eeta,\lambda)$ is included as supplementary material.

\bibliography{Ref_DPT} 
\bibliographystyle{apsrev}

\end{document}